\documentclass[twocolumn,pra,aps]{revtex4-1}

\usepackage[dvipdfmx]{graphicx}
\usepackage{color, fancybox} 
\usepackage{amsfonts}
\usepackage{amssymb,amsmath,amsthm} 
\usepackage{braket}

\begin{document}
\newcommand{\mbold}[1]{\mbox{\boldmath $#1$}}
\newcommand{\sbold}[1]{\mbox{\boldmath ${\scriptstyle #1}$}}
\newcommand{\tr}{\,{\rm tr}\,}
\renewcommand{\arraystretch}{1.5}

\newtheorem{theorem}{Theorem}
\newtheorem*{theorem*}{Theorem}
\newtheorem{lemma}{Lemma}
\newtheorem*{lemma*}{Lemma}

\title{Simple criterion for local distinguishability of generalized Bell states in prime dimension}
\author{T.~Hashimoto, M.~Horibe, and A.~Hayashi}
\affiliation{Department of Applied Physics, 
          University of Fukui, Fukui 910-8507, Japan}
\begin{abstract}
Local distinguishability of sets of generalized Bell states (GBSs) is 
investigated. We first clarify the conditions such that a set of GBSs  
can be locally transformed to a certain type of GBS set that is easily  distinguishable within local operations and one-way classical 
communication. We then show that, if the space dimension $d$ is a prime, 
these conditions are necessary and sufficient for sets of $d$ GBSs in  
$\mathbb{C}^d \otimes \mathbb{C}^d$ to be locally distinguishable. 
Thus we obtain a simple computable criterion 
for local distinguishability of sets of $d$ GBSs in prime dimension $d$.
\end{abstract}

\maketitle

\section{Introduction} 
\label{sec_introduction}
A set of orthogonal quantum states can be perfectly distinguished, though 
the laws of quantum mechanics do not allow one to distinguish 
non-orthogonal quantum states perfectly \cite{Helstrom76, Holevo82, Chefles00, Nielsen_text_book}. 
However, the issue is more involved when the states are shared by 
several parties and they are allowed to perform only local 
operations and classical communication (LOCC). Any two bipartite 
orthogonal states can be distinguished by one-way LOCC \cite{walgate2000}. For sets of three orthogonal states, however, some sets require 
two-way LOCC to distinguish, and some sets are not even locally 
distinguishable \cite{ghosh2001,nathanson2013}.     
In the space of $\mathbb{C}^d \otimes \mathbb{C}^d$, there are 
$d^2$ orthogonal states. 
However, it is impossible to locally 
distinguish more than $d$ orthogonal maximally entangled states 
perfectly in $\mathbb{C}^d \otimes \mathbb{C}^d$  
\cite{ghosh2004,nathanson2005,owari2006}. 

The generalized Bell states (GBSs) are typical examples of  
orthogonal maximally entangled 
states in $\mathbb{C}^d \otimes \mathbb{C}^d$, and 
much attention has been paid to clarifying local distinguishability 
of sets of GBSs \cite{fan2004,tian2016,singal2017,wang2017,yangwang2018,
yangyuan2018,wu2018,wang2019}. 
In particular some sufficient conditions for GBS sets to be one-way LOCC 
distinguishable or indistinguishable have been discussed in \cite{fan2004,wang2019,yangwang2018}. 

In this paper we present a simple computable criterion for 
local distinguishability (not limited to be one-way) of sets of 
$d$ GBSs in prime dimension $d$. In Sec.~\ref{sec_local_unitary} we  
clarify the conditions such that a set of GBSs can be 
locally transformed to a certain type of GBS set that can be distinguished 
by a simple strategy within one-way LOCC. In Sec.~\ref{sec_equivalence}, 
if $d$ is a prime, these conditions are shown to be necessary and 
sufficient for sets of $d$ GBSs in $\mathbb{C}^d \otimes \mathbb{C}^d$ 
to be locally distinguishable. Discussion including the case of  
composite-number dimensions are given in Sec.~\ref{sec_discussion}.

\section{Generalized Bell states}
\label{sec_GBS} 
Consider bipartite pure states shared by Alice and Bob in space 
$\mathbb{C}^d \otimes \mathbb{C}^d$. 
It is convenient to represent a maximally entangled state (MES) by 
a unitary operator $W$ in $\mathbb{C}^d$ 
(see, e.g., \cite{ghosh2004,nathanson2005,owari2006,fan2004}) as 
\begin{align}
   \ket{W}^{AB} &=  \frac{1}{\sqrt{d}}\sum_{a,b=0}^{d-1}W_{ab} 
                     \ket{a}^A\otimes\ket{b}^B
              \nonumber \\ 
   & = \left( W \otimes \mbold{1} \right) \ket{\mbold{1}}^{AB},
\end{align}
where 
\begin{align} 
    \ket{\mbold{1}}^{AB} =  \frac{1}{\sqrt{d}} 
                  \sum_{a=0}^{d-1} \ket{a}^A\otimes \ket{a}^B, 
\end{align}
with $\{ \ket{a} \}_{a=0}^{d-1}$ being an orthonormal base of 
$\mathbb{C}^d$.  
We note that $W$ denotes both the bipartite 
state and the unitary operator in this useful notation.  

The inner product between two MES states $\ket{W_1}$ and $\ket{W_2}$, 
expressed in terms of the corresponding operators $W_1$ and $W_2$, 
is given by 
\begin{align}
  \braket{W_1 | W_2} = \frac{1}{d} \tr{W_1^\dagger W_2}. 
        \label{eq_inner}
\end{align}
Suppose Alice and Bob perform some local unitary operations given by 
$A$ and $B$, respectively. We find that a MES state $\ket{W}$ is 
transformed to another MES state $\ket{AWB^T}$ as   
\begin{align}
  (A \otimes B) \ket{W} = \ket{AWB^T},
\end{align}
where the superscript $T$ represents the transposition with respect to 
the base $\{ \ket{a} \}_{a=0}^{d-1}$. 

Generalized Bell states (GBSs) belong to a special class of MESs, where 
the unitary operator $W$ is given by  
\begin{align}
   W_{m,n} = X^m Z^n,\ (m,n = 0,1,\ldots,d-1).  \label{eq_Wmn}
\end{align}
Here $X$ and $Z$ are generalized Pauli operators defined as 
\begin{align}
   X &= \sum_{a=0}^{d-1} \ket{a+1}\bra{a}, \\
   Z &= \sum_{a=0}^{d-1} \omega_d^a \ket{a}\bra{a},
\end{align}
where we employ the periodic convention for the base states; that is, 
$\ket{d+a}=\ket{a}$, and $\omega_d$ is a $d$th primitive root of unity, 
$\omega_d = e^{2\pi i/d}$. 
It is readily checked that the unitary operators 
$X$ and $Z$ satisfy the following relations:
\begin{align}
 X^d = Z^d = \mbold{1},\ ZX = \omega_d XZ.  \label{eq_XZrelation}
\end{align}
We can also see that the set of $d^2$ unitaries 
$\{ W_{m,n} \}_{m,n=0}^{d-1}$ in Eq.~(\ref{eq_Wmn}) is an orthonormal 
base in the operator space of $\mathbb{C}^d$. 
\begin{align}
  \tr{W_{m,n}^\dagger W_{m',n'}} = d \delta_{mm'}\delta_{nn'}.
\end{align}    
This together with Eq.~(\ref{eq_inner}) implies that $d^2$ GBSs 
$\{ \ket{W_{m,n}} \}_{m,n=0}^{d-1} $ 
form an orthonormal base in $\mathbb{C}^d \otimes \mathbb{C}^d$
, and therefore they are perfectly distinguishable by global measurements 
in the total space $\mathbb{C}^d \otimes \mathbb{C}^d$. 
However, it is known that Alice and Bob cannot distinguish more than  
$d$ MESs in $d \times d$ dimensions, if they are restricted to employ 
local operations and classical communication (LOCC) 
\cite{ghosh2004,nathanson2005,owari2006}.  

Suppose we are given a set of $\ell$ GBSs 
${\cal W} = \{ \ket{W_{m_i,n_i}} \}_{i=1}^\ell$ with $\ell \le d$.  
To specify a GBS set we will use the following notations interchangeably:
\begin{align}
 {\cal W} = \{ \ket{W_{m_i,n_i}} \}_{i=1}^\ell 
          = \{ W_{m_i,n_i} \}_{i=1}^\ell
          = \{ (m_i,n_i) \}_{i=1}^\ell.
\end{align} 
Our concern in this paper is what the conditions are for the set 
${\cal W}$ to be distinguishable by LOCC. 

Fan \cite{fan2004} noted that there is a special class of GBS sets 
for which one can easily find the way to distinguish the states 
with one-way LOCC. 
This is when all $m_i (i=1,2,\ldots,\ell)$ are distinct, 
and GBS sets with this property will be called F-type sets in this paper. 
Assume the set ${\cal W}$ is F type.  The states in the set 
are explicitly given by
\begin{align}
  \ket{W_{m_i,n_i}}  
      = \frac{1}{\sqrt{d}} \sum_{a=0}^{d-1} \omega_d^{n_i a} 
                    \ket{a+m_i}\ket{a},\ (i=1,\ldots,\ell).
\end{align}
Suppose Alice and Bob locally perform the projective measurement in the 
base $\{ \ket{a} \}_{a=0}^{d-1}$ and compare their outcomes. 
Then they obtain $m_i$, thereby identifying $i$ since all $m_i$ are 
distinct.

\section{Local unitary transformations and F-equivalent set}
\label{sec_local_unitary} 
Suppose a GBS set ${\cal W} = \{ (m_i,n_i) \}_{i=1}^\ell $ is 
transformed to another set  
${\cal W}' = \{ (m_i',n_i') \}_{i=1}^\ell$ by some  
local unitary transformations, 
\begin{align}
  U \otimes V \ket{X^{m_i}Z^{n_i}} &= \ket{ UX^{m_i}Z^{n_i}V^T } 
                \nonumber \\ 
   & \sim \ket{X^{m_i'}Z^{n_i'}}\ (i=1,\ldots,\ell),
\end{align}
where the symbol ``$\sim$'' means equality up to a global phase. 
It is clear that local distinguishability is invariant under 
local unitary transformations. 
If a set ${\cal W}$ can be transformed to a F-type set ${\cal W}'$ (defined in the preceding section), the set ${\cal W}$ is also distinguishable 
by one-way LOCC \cite{fan2004}. 

Let us determine the most general form of local unitary operations 
that transform all GBS states to other GBS states, that is, 
\begin{align}
    UX^{m}Z^{n}V^T \sim X^{m'}Z^{n'}\  (m,n = 0,\ldots,d-1). 
            \label{eq_UVtransformation}
\end{align}
A GBS set ${\cal W}$ is called F equivalent if  ${\cal W}$ can be 
transformed to a F-type set by this local unitary operation. 
By setting $m=n=0$ in Eq.~(\ref{eq_UVtransformation}) we have 
\begin{align}
  UV^T \sim X^{\mu_0}Z^{\nu_0},   \label{eq_UVTform}
\end{align}
for some integers $0 \le \mu_0,\nu_0 \le d-1$. 
This implies 
\begin{align}
  UX^{m}Z^{n}U^\dagger \sim X^{m''}Z^{n''}, \label{eq_Utransformation}
\end{align}
for some $m'',n''$, since
\begin{align*}
    & UX^{m}Z^{n}U^\dagger = UX^{m}Z^{n}V^T (UV^T)^{-1} 
               \nonumber \\
    & =  X^{m'}Z^{n'}(X^{\mu_0}Z^{\nu_0})^{-1} 
         \sim X^{m'-\mu_0}Z^{n'-\nu_0}. 
\end{align*}
Some specific types of unitary operators $U$ with this property 
have been used to study the F equivalence of GBS sets 
\cite{fan2004,wang2019}.  In this paper, we will employ more general 
unitary operators by establishing the necessary and sufficient conditions 
for the existence of unitary $U$ with the 
property of Eq.~(\ref{eq_Utransformation}).

First assume that Eq.~(\ref{eq_Utransformation}) holds for 
some unitary $U$. 
Setting $m=1,n=0$ or $m=0,n=1$, we obtain 
\begin{align}
  \left\{\begin{array}{ll}
   X' &  \equiv UXU^\dagger \sim X^\alpha Z^\gamma,  \\
   Z' &  \equiv UZU^\dagger \sim X^\beta Z^\delta,
         \end{array} 
  \right.
\end{align}
for some integers $0 \le \alpha,\beta,\gamma,\delta \le d-1$. 
It is clear that the relations given by Eq.~(\ref{eq_XZrelation}) persist 
under a unitary transformation, implying 
$X'^d=Z'^d=\mbold{1}$ and $Z'X' = \omega_d X'Z'$.  From the latter 
relation we find that $\alpha,\beta,\gamma,\delta$ should satisfy 
\begin{align}
 \det \left( \begin{array}{cc} 
            \alpha  & \beta \\
            \gamma  & \delta 
         \end{array}
      \right)
 \equiv 1 \pmod{d}.   \label{eq_sp1}
\end{align}

Conversely, assume that integers 
$0 \le \alpha,\beta,\gamma,\delta \le d-1$ satisfy the relation 
(\ref{eq_sp1}), and define 
\begin{align}
  \left\{\begin{array}{ll}
   X' &  \sim X^\alpha Z^\gamma, \\
   Z' &  \sim X^\beta Z^\delta. 
         \end{array} 
  \right.
\end{align}
We have the relation $Z'X' = \omega_d X'Z'$ and, furthermore, we can 
evidently choose global phase factors of $X',Z'$ such that 
$X'^d = Z'^d = \mbold{1}$ 
since $(X^\alpha Z^\gamma)^d \sim (X^\beta Z^\delta)^d \sim \mbold{1}$. 
We can show that the set of operators $\{X',Z' \}$ is unitary 
equivalent to $\{X,Z \}$; that is, there exists a unitary operator 
such that $ X' = UXU^\dagger, Z' = UZU^\dagger $. 
This can be seen in the following way: Let us take an eigenstate $\ket{\psi_0}$ of $Z'$ with an eigenvalue being a $d$th root of 
unity, $\omega_d^{k_0}$. 
Using the relation $Z'X' = \omega_d X'Z'$ repeatedly, we find that 
$ \ket{\psi_k} \equiv X'^k \ket{\psi_0}$ is an eigenstate of $Z'$ with 
eigenvalue $\omega_d^{k_0+k}$. Now it is clear that the relations 
$ X' = UXU^\dagger, Z' = UZU^\dagger $ hold if we take 
$U = \sum_{k=0}^{d-1} \ket{\psi_{k-k_0}}\bra{k}$. 
 
Thus we establish the following lemma: 
\begin{lemma} There exists a unitary operator $U$ such that 
\begin{align}
  \left\{\begin{array}{ll}
   UXU^\dagger \sim X^\alpha Z^\gamma,  \\
   UZU^\dagger \sim X^\beta Z^\delta,  \label{eq_lemma1-1}
         \end{array} 
  \right.
\end{align}
if and only if integers  $0 \le \alpha,\beta,\gamma,\delta \le d-1$ satisfy
\begin{align}
 \det \left( \begin{array}{cc} 
            \alpha  & \beta \\
            \gamma  & \delta 
         \end{array}
      \right)
 \equiv 1 \pmod{d}.  \label{eq_lemma1-2}
\end{align}
\end{lemma}

All $2 \times 2$ matrices 
$\left(\begin{smallmatrix}
               \alpha & \beta \\ 
               \gamma & \delta
       \end{smallmatrix}
\right)$ with integer entries satisfying 
$ 0 \le \alpha,\beta,\gamma,\delta \le d-1$ and 
the condition given by Eq.~(\ref{eq_lemma1-2}) form a group under matrix 
multiplication modulo $d$. This group is denoted by $Sp(d)$.  
We note that $U$ is not unique for a given element of $Sp(d)$. 
It depends on global phase factors that are not specified in 
Eq.~(\ref{eq_lemma1-1}). The phase of 
$U$ itself is not determined either since the unitary 
transformations of Eq.~(\ref{eq_lemma1-1}) are independent of a  
phase change, $U \rightarrow e^{i\theta} U$. 
For discussions on explicit forms of $U$ 
corresponding to a given element of $Sp(d)$, see Ref.~\cite{watanabe2019}, where a Euclidean type of algorithm to construct 
$U$ is presented. 

Using $ UX^mZ^nV^T = UX^mZ^nU^\dag(UV^T) $
together with  Eqs.~(\ref{eq_UVTform}) and (\ref{eq_lemma1-1}), 
we conclude that operator $X^mZ^n$ transforms under local transformations 
as 
\begin{align}
  UX^mZ^nV^T  \sim X^{m'}Z^{n'}, 
\end{align}
where 
\begin{align}
  \left( \begin{array}{c}
             m' \\
             n' 
         \end{array}
  \right) 
  \equiv 
  \left( \begin{array}{cc} 
            \alpha  & \beta \\
            \gamma  & \delta 
         \end{array}
  \right)
  \left( \begin{array}{c}
             m \\
             n 
         \end{array}
  \right) 
  +
  \left( \begin{array}{c}
             \mu_0 \\
             \nu_0
         \end{array}
  \right) 
   \pmod{d}.     \label{eq_spdTransformation}
\end{align}

Now it is easy to write the conditions for a GBS set 
${\cal W} = \{ (m_i,n_i) \}_{i=1}^\ell $ to be F equivalent. 
The conditions are that 
\begin{align}
      m_i' = m_i \alpha + n_i \beta\ + \mu_0\ (i=1,\ldots,\ell),
\end{align} 
are all distinct modulo $d$ for some $Sp(d)$ matrix 
$\left(\begin{smallmatrix}
               \alpha & \beta \\ 
               \gamma & \delta
       \end{smallmatrix}
\right)$ and an integer $0 \le \mu_0 \le d-1$. 

It is clear that the integer $\mu_0$ can be omitted in the above 
conditions.  
As for $\alpha$ and $\beta$, it is assumed 
that they are the elements of some $Sp(d)$ matrix  
$\left(\begin{smallmatrix}
               \alpha & \beta \\ 
               \gamma & \delta
       \end{smallmatrix}
\right)$. 
However, these constraints can be lifted; that is, $\alpha$ and 
$\beta$ are any integers. This can be seen in the following way: 
Suppose that $m_i \alpha + n_i \beta$ are all distinct modulo $d$ for 
some integers $\alpha$ and $\beta$. It is clear that 
$m_i \alpha_1 + n_i \beta_1$ are also all distinct modulo $d$, where 
$\alpha_1 \equiv \alpha, \beta_1 \equiv \beta \pmod{d}$ and 
$ 0 \le \alpha_1,\beta_1 \le d-1$. Write $\alpha_1 = c\alpha_2, 
\beta_1 = c\beta_2$ with $c = \gcd(\alpha_1,\beta_1)$. Note that 
$\gcd(\alpha_2,\beta_2)=1$.
Then we find that $m_i \alpha_2 + n_i \beta_2$ are all distinct 
modulo $d$ and there are some integers $0 \le \gamma_2,\delta_2 \le d-1$ 
such that $\alpha_2\delta_2-\beta_2\gamma_2 \equiv 1 \pmod{d}$.

Thus we arrive at the following theorem:
\begin{theorem}
A set of $\ell$ GBSs ${\cal W}$ in 
$\mathbb{C}^d \otimes \mathbb{C}^d$ is F equivalent and therefore 
one-way LOCC distinguishable if and only if 
\begin{align}
     m_i \alpha + n_i \beta\  (i=1,\ldots,\ell),
\end{align} 
are all distinct modulo $d$ for some integers $\alpha$ and $\beta$. 
\end{theorem}

Theorem 1 gives a sufficient condition for a GBS set ${\cal W}$ to be 
distinguishable by one-way LOCC. We show that the same condition 
can be derived from a different point of view. 
Ghosh {\it et al.} showed that a GBS set ${\cal W}$ is one-way LOCC  distinguishable if and only if there is some state $\ket{\phi}$ 
such that $\{ X^{m_i}Z^{n_i} \ket{\phi} \}_i^\ell$ are pairwise 
orthogonal \cite{ghosh2004}.
This orthogonality is expressed as 
\begin{align}
   \braket{\phi | X^{m_i-m_j}Z^{n_i-n_j} |\phi} = 0,\ i \ne j. 
     \label{eq_ghosh}
\end{align} 
We show that in some cases, the state $\ket{\phi}$ satisfying this equation 
can easily be found. 
To do so, we will employ a general property of 
two unitary operators. Let $V$ and $V'$ be unitary, and assume that 
$VV'=\lambda V'V$ with $\lambda \ne 1$. 
For any eigenstate $\ket{\phi}$ of $V$,  we find 
\begin{align}
 \braket{\phi|V'|\phi} = \lambda\braket{\phi|V^\dagger V'V|\phi}
  = \lambda\braket{\phi|V'|\phi}.
\end{align} 
From this, it follows that $\braket{\phi|V'|\phi}=0$ since $\lambda \ne 1$.

Now let $V$ be  $X^{-\beta} Z^\alpha$ with some integers $\alpha,\beta$
and suppose that $V$ does not commute with $V' = X^{m_i-m_j}Z^{n_i-n_j}$ 
for every $i \ne j$. This noncommutability 
can be expressed as the following conditions:
\begin{align}
   (m_i-m_j)\alpha + (n_i-n_j)\beta \ne 0 \pmod{d},\ i \ne j, 
\end{align}
which is equivalent to the conditions in Theorem 1. 
Taking $\ket{\phi}$ to be an eigenstate $V$, we obtain   
Eq.~(\ref{eq_ghosh}), which shows that the set ${\cal W}$ is one-way 
LOCC distinguishable.

\section{Necessary and sufficient condition for local distinguishability 
in the case of prime $\ell=d$}
\label{sec_equivalence}
Theorem 1 in the preceding section gives a sufficient condition 
for a GBS set ${\cal W}=\{ \ket{W_{m_i,n_i}} \}_{i=1}^\ell$ 
in $\mathbb{C}^d \otimes \mathbb{C}^d$ to be distinguishable by 
one-way LOCC for arbitrary $\ell$ and $d$. 
In this section, we consider the case where $d$ is a prime number and 
$\ell=d$. Then it will be shown that the condition given in Theorem 1 
is also necessary for one-way LOCC distinguishability; F equivalence is equivalent to one-way LOCC distinguishability. 
Furthermore we will see that the restriction to one-way LOCC can be 
removed by using a lemma given by Yu and Oh \cite{yu2015}. 

Let us assume that a set of $d$ GBSs 
${\cal W}=\{ \ket{W_{m_i,n_i}} \}_{i=1}^d$ 
is one-way LOCC distinguishable and show that the set ${\cal W}$ is 
then F equivalent.  According to Ghosh {\it et al.}, 
there is a normalized state $\ket{\phi}$ such that
\begin{align}
      \braket{\phi | W_{m_i,n_i}^\dagger W_{m_j,n_j} |\phi} 
       = \delta_{ij}\ (i,j=1,\ldots,d), 
             \label{eq_orthogonality}
\end{align}
which implies 
\begin{align}
 \sum_{i=1}^d  W_{m_i,n_i} \ket{\phi}\bra{\phi}W_{m_i,n_i}^\dagger
               = \mbold{1}.
             \label{eq_completeness}
\end{align} 
For $\ket{\phi}\bra{\phi}$ on the left-hand side, we substitute its  expanded form in terms of the complete operator set 
$\{ W_{m,n} \}_{m,n=0}^{d-1}$, 
\begin{align}
  \ket{\phi}\bra{\phi} = \frac{1}{d} \sum_{m,n=0}^{d-1} 
        \braket{\phi| W_{m,n}^\dagger |\phi} W_{m,n}. 
\end{align}
We obtain 
\begin{align}
  \sum_{m,n=0}^{d-1} 
        \braket{\phi| W_{m,n}^\dagger |\phi} 
            \kappa_{mn}  W_{m,n} = \mbold{1}, \label{eq_1expansion}
\end{align}
where $\kappa_{mn}$ is defined as 
\begin{align}
  \kappa_{mn} = \frac{1}{d} \sum_{i=1}^{d} \omega_d^{n_im-m_in}.
\end{align}
Note that Eq.~(\ref{eq_1expansion}) is the expansion form 
of the identity $\mbold{1}$ in terms of $W_{m,n}$.  
When $(m,n) \ne (0,0)$, the coefficients 
$\braket{\phi| W_{m,n}^\dagger |\phi} \kappa_{mn}$ should vanish. 
Evidently there are some $(m,n) \ne (0,0)$ such that
$\braket{\phi| W_{m,n}^\dagger |\phi} \ne 0$, which requires  
$\kappa_{mn}=0$.  

Here we employ the following lemma:
\begin{lemma}
Let $\omega_d = e^{2\pi i/d}$, with $d$ being a prime. 
Assume 
\begin{align}
  \sum_{i=1}^d \omega_d^{\nu_i} = 0, 
\end{align}
for some $d$ integers $0 \le \nu_i \le d-1$.  
This is possible if and only if all $\nu_i$ are distinct, i.e., 
$\{\nu_i\}_{i=1}^d = \{0,1,\ldots,d-1\}$. 
\end{lemma}

The proof of Lemma 2 will be given at the end of this section.  
We have shown that if the GBS set ${\cal W}$ is one-way distinguishable, 
then $\kappa_{m,n} = 0$ for some integers $m,n$. 
According to Lemma 2, this implies that 
$n_i m -m_i n\ (i=1,\ldots,d)$ are distinct modulo $d$ 
for some integers $m,n$, and therefore, by Theorem 1, we conclude that 
the set ${\cal W}$ is F equivalent.
 
Thus we have shown that a set of $d$ GBSs ${\cal W}$ in prime 
dimension is one-way LOCC distinguishable if and only if ${\cal W}$ is 
F-equivalent. 

As shown in the following, the restriction ``one way'' can 
actually be removed. For that, we employ the lemma of 
Yu and Oh \cite{yu2015}.
\begin{lemma*}[Yu and Oh]
Assume that  a set of $d$ GBSs ${\cal W} = \{W_{m_i,n_i}\}_{i=1}^d$ in 
$\mathbb{C}^d \otimes \mathbb{C}^d$ satisfies 
the following conditions: 
If $\sum_{i=1}^d \omega_d^{mn_i-nm_i} = 0$ 
, then $(m,n) \equiv (m_i-m_j,n_i-n_j) \pmod{d}$ for some 
$ i \ne j$. This is possible only when ${\cal W}$ is not 
distinguishable by LOCC. 
\end{lemma*}

Here LOCC is not restricted to be one way. 
This lemma was derived by a method 
of detecting the local indistinguishability proposed by
Horodecki {\it et al}. \cite{horodecki2003}. It is based on 
the fact that the LOCC transition of bipartite states
$\ket{\psi} \rightarrow \{p_i,\ket{\psi_i} \}$ is possible 
if and only if the vector $\sum_i p_i \lambda(\psi_i)$ majorizes 
$\lambda(\psi)$ \cite{jonathan1999}, where $\lambda$ is the vector 
of squared Schmidt coefficients. This is the reason why LOCC in 
this lemma is not restricted to be one way. 

Suppose that the set ${\cal W}$ is not F equivalent. As shown before, 
this implies that there are no integers $m,n$ such that 
$\sum_{i=1}^d \omega_d^{mn_i-nm_i} = 0$. The lemma of Yu and Oh tells 
that the set ${\cal W}$ is not distinguishable by LOCC. 
We thus obtain the main result of this paper. 
\begin{theorem}
A $d$-GBS set ${\cal W} = \{W_{m_i,n_i}\}_{i=1}^d$ 
in $\mathbb{C}^d \otimes \mathbb{C}^d$ with $d$ being prime is distinguishable by LOCC if and only if 
$W$ is F equivalent; that is, 
\begin{align}
     m_i \alpha + n_i \beta\  (i=1,\ldots,d),
\end{align} 
are all distinct modulo $d$ for some integers $\alpha$ and $\beta$. 
\end{theorem}
 
The rest of this section is devoted to the proof of Lemma 2. 
In the complex plane the points $\{\omega_d^\nu\}_{\nu=0}^{d-1}$ are 
at the vertices of a regular $d$-sided polygon inscribed in the unit 
circle.   The ``if'' part of the lemma is evident. 
For small primes ($d$ = 2,3), the ``only if'' part also appears to 
be evident. 
For larger primes, however, some knowledge of the cyclotomic polynomials 
is needed. 

The $n$th cyclotomic polynomial is defined to be 
\begin{align}
 \Phi_n(x) \equiv \prod_{\substack{1 \le \nu \le n \\ \gcd(\nu,n)=1}} 
                \left(x - e^{\frac{2\pi i}{n} \nu} \right).
\end{align}
Its roots are all $n$th primitive roots of unity. It can be shown
that the coefficients of the cyclotomic polynomials are integers. 
For example, we find 
\begin{align}
  &\Phi_1(x) = x-1,\ \Phi_2(x) = x+1,\ \Phi_3(x) = x^2+x+1,
           \nonumber \\ 
  &\Phi_4(x) = x^2+1,\ \Phi_5(x) = x^4+x^3+x^2+x+1,
           \nonumber \\ 
  &\Phi_6(x) = x^2-x+1,\cdots .
\end{align} 
For a prime $n$, $\Phi_n(x)$ is clearly given by
\begin{align}
 \Phi_n(x) = \frac{x^n-1}{x-1} = \sum_{\nu=0}^{n-1} x^\nu,
\end{align}
since all $n$th roots of unity are primitive except for unity itself. 
One of the remarkable properties of the cyclotomic polynomials is 
that $\Phi_n(x)$ is irreducible over $\mathbb{Q}[x]$ (all polynomials 
with rational coefficients)  \cite{Schroeder_text_book,Lang_text_book}.  
It has no nontrivial factors in $\mathbb{Q}[x]$ with smaller 
degree, and therefore it is the unique minimal polynomial of 
$e^{\frac{2\pi i}{n}}$ over $\mathbb{Q}[x]$.
This means that if a polynomial $f(x)$ in $\mathbb{Q}[x]$ is monic
(the leading coefficient is 1) and it satisfies $f(e^{\frac{2\pi i}{n}})=0$, 
then we have $\deg f(x) > \deg \Phi_n(x)$ or $f(x)=\Phi_n(x)$.    

Suppose that the relation  $\sum_{i=1}^d \omega_d^{\nu_i} = 0$ holds,  
and consider the following polynomial of $x$: 
\begin{align}
  f_d(x) \equiv \frac{\sum_{i=1}^d x^{\nu_i}}
          {\text{the leading coefficient of }\sum_{i=1}^d x^{\nu_i}}.
\end{align}
We then observe  
\begin{enumerate}
\item $f_d(x)$ is a polynomial of $x$ with rational coefficients, and 
it is monic.
\item $f_d(\omega_d) = 0$.
\item $\deg f_d(x) \le d-1$.
\end{enumerate}
Since $\deg \Phi_d(x) = d-1$ for a prime $d$, we conclude 
$f_d(x) = \Phi_d(x)$, which is possible only if all $\nu_i$ are distinct.
This completes the proof of Lemma 2.

\section{Discussion and concluding remarks}
\label{sec_discussion}
We have shown that local distinguishability is equivalent to 
F equivalence for a set of $d$ GBSs in $\mathbb{C}^d\otimes \mathbb{C}^d$ 
with prime $d$. Here it should be emphasized that the GBS set that cannot 
be transformed to be F type is not distinguishable even with two-way 
LOCC. Theorems 1 and 2 provide a computable simple criterion for that: 
a finite number of integer calculations are sufficient to test whether 
a GBS set is F equivalent. 

It is not straightforward to extend this conclusion to general 
$\ell < d$ cases. One reason for this can be seen in the 
rewriting of ``orthogonality'' in Eq.~(\ref{eq_orthogonality}) to ``completeness'' in Eq.~(\ref{eq_completeness}), 
which is possible only if the number of states is equal to the space 
dimension. 

Let us take some cases where the dimension $d$ is not prime.
Consider $d$-GBS sets with $d=d_1^2$.  In the case of $d=4$, 
there are two types of one-way LOCC distinguishable GBS sets that are not 
F equivalent \cite{tian2016,singal2017}. 
One of them is ${\cal W} = \{(0,0),(0,2),(2,0),(2,2) \}$. 
This set can easily be generalized to general $d=d_1^2$ cases. 
Consider the GBS set given by 
\begin{align}
  {\cal W} = \{ X^{\mu d_1}Z^{\nu d_1} \}_{\mu,\nu=0}^{d_1-1}. 
\end{align}
Clearly the set ${\cal W}$ is not F type. It is not even F equivalent since 
it is invariant under the $Sp(d)$ 
transformations given in Eq.~(\ref{eq_spdTransformation}).
However, the set ${\cal W}$ is one-way LOCC distinguishable. 
To see this, take 
\begin{align}
   \ket{\phi} = \frac{1}{\sqrt{d_1}}\, 
               ( \overbrace{1,1, \ldots ,1}^{d_1},
                 \overbrace{0,0, \ldots \ldots \ldots ,0}^{d_1^2-d_1}
               ).
\end{align}
Then we find that $d$ states given by
\begin{align}
     X^{\mu d_1}Z^{\nu d_1} \ket{\phi}= 0,\ \mu,\nu=0,\ldots,d_1-1, 
\end{align}
are pairwise orthogonal, showing that ${\cal W}$ is one-way LOCC 
distinguishable.
When $d=d_1^2$, we thus found that distinguishability by one-way LOCC 
$\supsetneq$ F equivalence.

We performed some numerical analysis to test distinguishability 
by one-way LOCC for all sets of six GBSs in $6 \times 6$ dimension, 
which is the simplest example for $d=d_1d_2$ with relatively prime 
$d_1$ and $d_2$.  
The results indicate that distinguishability by one-way LOCC 
is equivalent to F equivalence in this example. 
Further studies are needed in order to clarify 
how this equivalence persists in general $d=d_1d_2$ cases.   


\end{document}